\documentclass[letterpaper,10pt,onecolumn]{article}

\usepackage{lmodern}

\usepackage{color}
\usepackage{amsmath,amsfonts,amssymb,mathtools}
\usepackage{graphicx}

\usepackage[small]{titlesec}

\usepackage{fullpage}
\usepackage[pass,includehead,includeheadfoot,showframe=false]{geometry}

\usepackage[mathlines]{lineno}
\modulolinenumbers[2]

\usepackage{setspace}
\def\Blue#1{#1}

\usepackage{xspace}
\newcommand{\icon}[0]{ICON\xspace}	

\usepackage[font=small]{caption}

\begin{document}

\begin{center}
{\Large\bf \icon: an adaptation of infinite HMMs for time traces with drift}
\\[2ex]
\large I Sgouralis and S Press{\'e}
\end{center}

\begin{abstract}
Bayesian nonparametric methods have recently transformed emerging areas within data science. 
One such promising method, the infinite hidden Markov model (iHMM), generalizes the 
HMM which itself has become a workhorse in single molecule data analysis.
The iHMM goes beyond the HMM by 
\Blue{self-consistently learning all parameters learned by the HMM in addition to learning the number of states
without recourse to any model selection steps.
Despite its generality, simple features (such as drift), common to single molecule time traces,
result in an over-interpretation of drift and the introduction of artifact states. Here we present an adaptation of the iHMM that can treat data with drift originating from one or many traces (e.g. FRET).} 
Our fully Bayesian method couples the iHMM to a continuous control process (drift)
self-consistently learned while learning all other quantities determined by the iHMM \Blue{(including state numbers).
A key advantage of this method is that all traces -- regardless of drift or states visited across traces --
may now be treated on an equal footing thereby eliminating user-dependent trace selection 
(based on drift levels), pre-processing to remove drift and post-processing model selection on state number.} 
\end{abstract}

\section{Introduction}

Single molecule experiments -- such as single molecule FRET (smFRET) \cite{roy2008practical} or force spectroscopy~\cite{neuman2008single} -- 
exhibit discrete transitions between states or molecular conformations. 
These transitions are often idealized as memoryless (Markovian) processes and, 
as a result, hidden Markov models (HMM)  \cite{rabiner1986introduction} -- describing Markov transitions obscured by noise --
naturally arise in single molecule data analysis~\cite{bronson2009learning,blanco2010analysis,yoon2008bayesian,rosales2004mcmc,hines2015primer,mckinney2006analysis}. 
While traditional HMMs are used to find kinetics of transition between states, they cannot learn the number of states. Indeed, for the HMM to apply in the first place, 
the number of states that the system may visit over the course of a time trace must be fixed {\it a priori} \cite{blanco2010analysis}. 

Following a HMM analysis, model selection criteria are often invoked to discriminate between
various models -- labeled ``parametric'' because they entail a \emph{finite} set of parameters --
with a different number of states~\cite{mckinney2006analysis,munro2007identification,bronson2009learning}. This approach breaks down the full optimization problem of i) enumerating the states and ii) parametrizing the model into two disjoint operations that may not result in an optimal global solution.

It is for this reason that a nonparametric~\cite{ferguson1983bayesian} realization of the HMM, the so called \emph{infinite hidden Markov model} (iHMM) was developed \cite{beal2001infinite}. Since its introduction, the iHMM has provided a principled method to perform time series analysis in machine learning applications. Despite the success of iHMMs outside Biology, to our knowledge only a single proof-of-concept paper illustrating the potential of iHMMs to single molecule 
Biophysics has appeared \cite{hines2015analyzing}.

By contrast to HMMs having a fixed number of states,
the nonparametric adaptation of the HMM, the iHMM, may recruit from an infinite pool
new states on the basis of the available data~\cite{beal2001infinite,teh2012hierarchical,ferguson1983bayesian}
without ever prespecifying a total number as would be the case with the HMM.

\begin{figure}[tbp]
   \centering
   \includegraphics[scale=1]{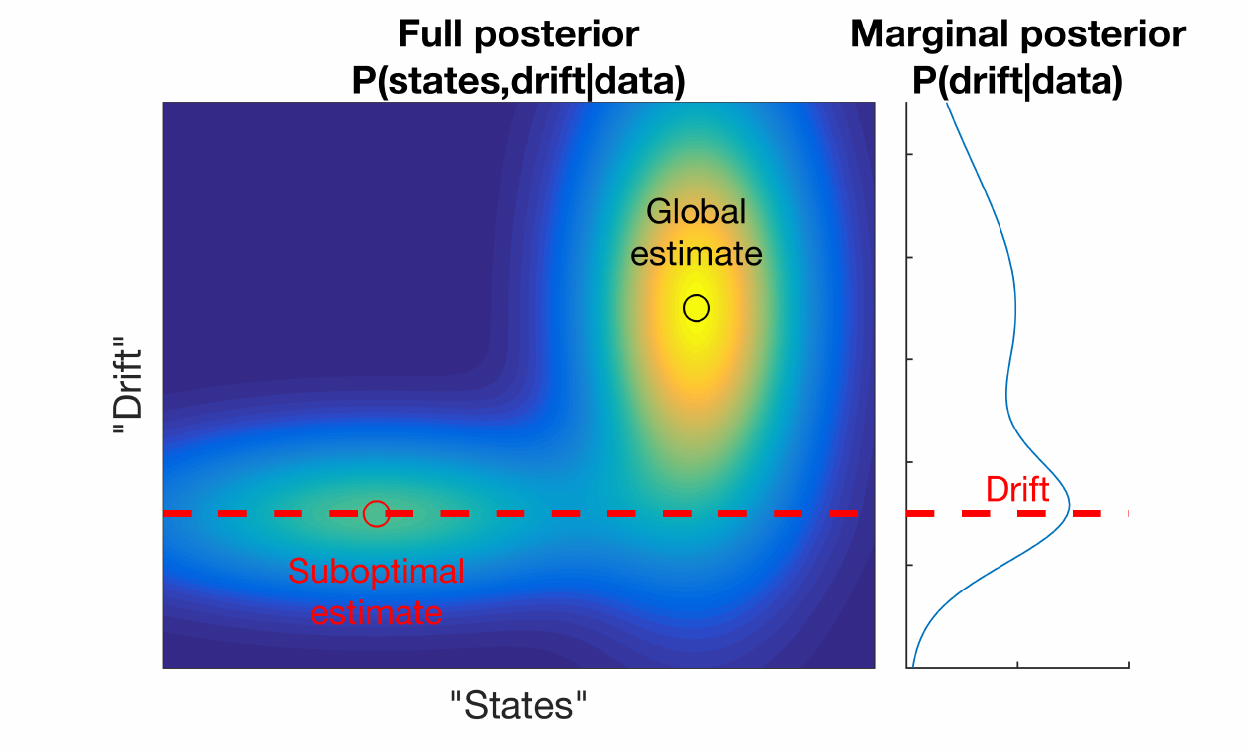} 
   \caption{\Blue{{\bf A conceptual illustration of the pitfalls of independent de-drifting in the analysis of single molecule measurements}. Here, the axes represent abstractions of a single molecule's states and drift. The model's posterior probability distribution (main panel) is color coded with brighter colors corresponding to higher probability. Inference primarily seeks to identify the most likely states estimate, i.e.\ the estimate with the highest posterior probability. In cases where the posterior has multiple maxima as shown, processing for drift independently of the states might lead to inaccurate estimation of drift (dashed line). Here, drift estimation is based on the posterior distribution marginalized over the states (left panel). Subsequently, processing of the states, which in this case will be restricted only along the dashed line, will result in less likely state number estimates. By contrast, simultaneous processing of the states and drift as we proposed herewith allow inference in the full plane. Hence the global maximum, i.e.\ the most likely combination of states and drift, can be readily identified. For an explicit example utilizing inaccurate drift estimates see lower panel of Fig.~\ref{fig:syn_1} below. Ignoring drift in this example is equivalent to misidentifying a flat drift profile and sub-optimality of the inferred states is demonstrated by the identification of artifact states.}}
   \label{fig:concept}
\end{figure}

While promising, the iHMM's flexibility -- lending the iHMM its intrinsic ability to learn the number of conformational states, say, visited by a single molecule -- can become an important weakness. As an example, iHMMs would deal with drift, encountered throughout single molecule experiments, by adding ``artifact states". 
To date, just as it is common to prespecify the number of states in an HMM, it is also common practice to de-drift (or de-trend) the time traces \Blue{using standard time series methods, e.g.\ \cite{douc2014nonlinear,wu2007trend,hamilton1994time},} and subsequently analyze the data with an HMM \cite{holden2010defining}. 
In principle, the same may be done with iHMMs.
The catch however is that, at best, de-drifting separately from the rest of the data analysis leads to a subsequent estimation of the states
conditioned on the de-drifting procedure which, in turn, may again
lead to suboptimal estimates for the states as conceptually illustrated in Fig.~\ref{fig:concept}. 

In fact, it is specifically to avoid such suboptimal estimates (in this case to estimate noise properties while estimating transition probabilities)
that HMMs exist in the first place: they learn transition probabilities while de-noising a time trace
self-consistently.

Here we take the same logic some steps further and adapt the iHMM to make it useful \Blue{for applications in Biophysics where traces are often corrupted by drift}. In particular, we provide \Blue{a novel formulation of iHMMs with a fully Bayesian consideration of drift}. We adapt the iHMM to: i) incorporate information from multiple traces (e.g.~recordings from different channels) -- typically available in many setups such as FRET measurements -- and; ii) infer the drift simultaneously, and thus self-consistently, while learning the number of states along with their transition kinetics and emission properties.

The remainder of this paper is organized as follows. In Methods, we present the formulation of the iHMM and its extensions leading to \icon (Infinite HMM coupled to a continuous CONtrol process, i.e.~the drift) which are required for the subsequent analysis. In the Results, we present selected applications of the proposed method using both synthetic and experimental datasets. Lastly, in the Discussion, we discuss the broader potential of the method to Biophysics. \Blue{Additional remarks and an implementation of the proposed method can be found in the supporting materials.}

\section{Methods}
\label{sec:methods}

\Blue{Here we first describe the basic formulation of the HMM as it applies to biophysical data allowing us to introduce the iHMM as its generalization. Subsequently, we introduce \icon as an extension of the iHMM.} While the literature on iHMMs is currently intended for specialized audiences in Statistics, and the mathematics themselves are quite technical~\cite{teh2012hierarchical}, 
we provide a more detailed presentation of the iHMM dedicated to physical scientists in a companion article~\cite{perspectives}. 

\begin{figure}[tbp]
   \centering
   \includegraphics[scale=1]{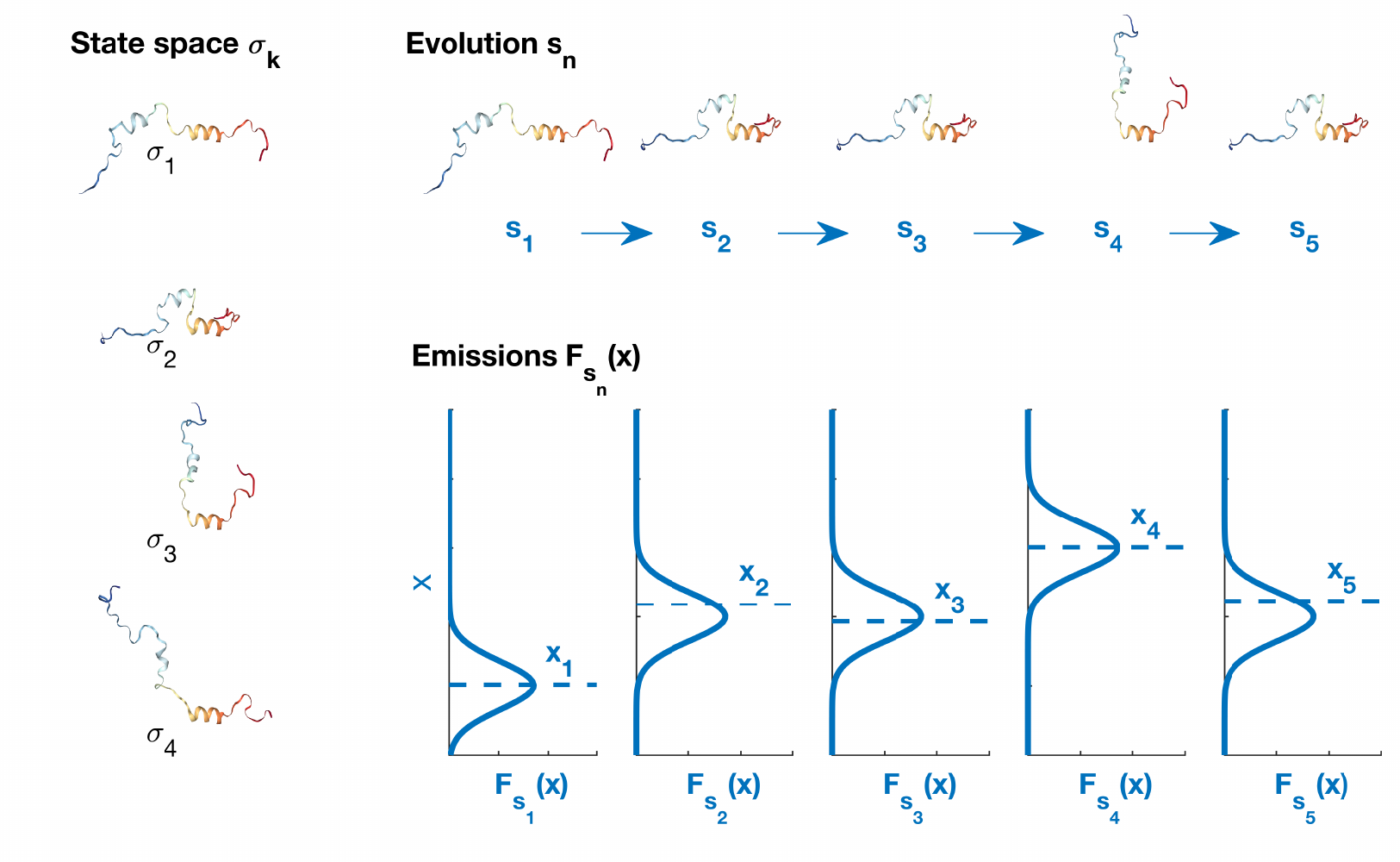} 
   \caption{{\bf Illustration of a hypothetical biomolecule with four or potentially more conformational states}. \emph{Left:}~the biomolecule's conformations $\sigma_k$. \emph{Upper right:} Sequence of conformational changes at each time step, $s_n$. For sake of illustration here, only conformations $\sigma_1$, $\sigma_2$, and $\sigma_3$ are visited, while $\sigma_4$ remains unvisited throughout the time trace. \emph{Lower right:}~Corresponding emission distributions $F_{s_n}(x_n)$ of the visited states -- since the system remains in the same state at time steps 2 and 3, those emission distributions are identical. 
Dashed lines represent the emissions $x_n$ at corresponding times $t_n$. For the biomolecule's illustration, we used data from Ref.~\cite{chong2016new}.}
   \label{fig:stsp}
\end{figure}

We begin by assuming that a molecule under investigation switches, on \Blue{or slower than} the data acquisition timescale, discretely between different conformational states that we denote $\sigma_k$, where $k=1,2,\dots$ labels these conformations, Fig~\ref{fig:stsp}. 
\Blue{While we will keep referring to single molecules for simplicity, 
it is understood that time traces similar to the ones we consider 
may also be generated from experiments on molecular complexes involving many biomolecules ~\cite{sen2013clpxp}}.

Each conformation $\sigma_k$ yields or ``emits" observations $x$ according to a unique probability distribution $F_{\sigma_k}(x)$. 
\Blue{To model each of these distributions, we assume a generic form that depends on some state specific parameters $\phi_{\sigma_k}$. In this study, for simplicity, we assume Gaussian emissions} 
\begin{linenomath*}\begin{align}\label{eq:F_emiss}
F_{\sigma_k}(x)=\sqrt{\frac{\tau_{\sigma_k}}{2\pi}}\exp\left(-\frac{\tau_{\sigma_k}}{2}(x-\mu_{\sigma_k})^2\right)
\end{align}\end{linenomath*}
\Blue{where, for this specific example, the parameters include the Gaussian mean and precision, $\phi_{\sigma_k}=(\mu_{\sigma_k},\tau_{\sigma_k})$.} In general, the method is readily customized to treat any family of emission distribution, such as Poisson, that may 
more accurately capture experimental conditions \cite{neal2000markov}.

During the experiment, emissions are collected at equidistant time intervals or ``steps'' $t_n$, where $n=1$ and $N$ denote the first and last measurements. At each time step, we denote the corresponding emission as $x_n$ and the underlying conformation as $s_n$. That is, during the experiment, the biomolecule choses $s_n$ from the set $\{\sigma_1,\sigma_2,\dots\}$ and emits $x_n$ according to $F_{s_n}(x_n)$; for an illustration see Fig.~\ref{fig:stsp}.

To model the sequence of the successive conformations $s_1\to s_2\to\cdots\to s_N$ that the biomolecule attains during the experiment time course, HMMs assume Markovian dynamics \cite{rabiner1986introduction,mckinney2006analysis} leading to the following scheme
\begin{linenomath*}\begin{align}
s_n\big| s_{n-1}&\sim Cat(\tilde\pi_{s_{n-1}}),	\label{eq:trans}
\\
x_n\big| s_n&\sim F_{s_n}.	\label{eq:emiss}
\end{align}\end{linenomath*}
Here, $\tilde\pi_{s_{n-1}}=(\pi_{s_{n-1}\to\sigma_1},\pi_{s_{n-1}\to\sigma_2},\dots)$ gathers the probabilities of switching from $s_{n-1}$ to any $\sigma_k$, and $Cat(\tilde\pi_{s_{n-1}})$ denotes the categorical probability distribution. In other words, 
$\pi_{s_{n-1}\to\sigma_k}$ is the probability of jumping from $s_{n-1}$ to $\sigma_k$. 

From the HMM model, Eqns.~\eqref{eq:trans}--\eqref{eq:emiss}, the logic proceeds forward as follows: we derive a likelihood (or a posterior probability) of observing the time trace
from the model and subsequently maximize this likelihood (or posterior probability) to find the values of the model parameters~\cite{rabiner1986introduction}. 
These parameters include the transition probabilities $\pi_{\sigma_k\to\sigma_j}$ between all pairs of states ($\sigma_k$ and $\sigma_j$) and all emission parameters 
$\phi_{\sigma_k}$.  

The iHMM is more complex than the HMM and relies on mathematics outside the scope of this paper
that we have however detailed in a companion article  \cite{perspectives}.

In the iHMM, the precise number of conformational states attainable by a biomolecule -- and thus the size of $\{\sigma_1,\sigma_2,\dots\}$ --  
is \emph{a priori} left unspecified \cite{beal2001infinite}. 
Starting from a conformation $s_n$ at time step $n$, the biomolecule is allowed to choose between a potentially infinite number of conformations $\sigma_k$ for the next time step. Only the form of the prior on those transition probabilities -- in other words, a probability distribution over $\tilde\pi_{\sigma_k}$ -- 
determines whether the system visits a new or an already visited state at each time step.
As explained in Ref.~\cite{teh2012hierarchical} and Ref.~\cite{perspectives}, this prior is a
\emph{hierarchal Dirichlet process} which, in one of its realizations, takes the form 
\begin{linenomath*}\begin{align}
\tilde\beta&\sim GEM(\gamma),
\\
\tilde\pi_{\sigma_k}\big|\tilde\beta&\sim DP(\alpha,\tilde\beta),
\\
\phi_{\sigma_k}&\sim H	\label{eq:phi},
\end{align}\end{linenomath*}
where $GEM(\gamma)$ and $DP(\alpha,\tilde\beta)$ denote a stick-breaking and a Dirichlet process with concentration parameters $\gamma$ and $\alpha$, respectively, and $\tilde\beta=(\beta_{\sigma_1},\beta_{\sigma_2},\dots)$ is the base distribution \cite{teh2012hierarchical}. Here, $H$ denotes the prior probability distribution of the emission parameters $\phi_{\sigma_k}$. Throughout this study, for $H$, we adopt the conditionally-conjugate model developed in Ref.~\cite{gorur2010dirichlet}, \Blue{which allows simultaneous learning of $\mu_{\sigma_k}$ and $\tau_{\sigma_k}$;} see~Eq.~\eqref{eq:F_emiss}. The full set of equations formulating iHMM is listed in the supporting materials.

The iHMM captures the fact that
measurements are contaminated with noise described by the state-dependent emission distributions $F_{\sigma_k}(x)$. However, perhaps even more importantly, time traces may also be contaminated with drift \cite{roy2008practical} which the model -- as written in Eqns.~\eqref{eq:trans}--\eqref{eq:phi} 
-- does not capture. 
This happens because iHMMs model abrupt transitions in traces which, for the moment, excludes the possibility of a slowly-evolving drift that requires a \emph{continuous} process instead.

To account for drift -- a necessary ingredient in adapting iHMMs to single molecule data analysis -- the emission distributions of Eq.~\eqref{eq:emiss} must be coupled to a continuous process that acts like a ``control'' or ``drive''. For simplicity, we call this the ``control process".

Below, we present the extensions needed to generalize the iHMM into \icon. 
Initially, we assume single trace observations (as would be the case if we were to look at a force spectroscopy time trace or a single channel from an smFRET experiment)
and describe how we account for drift. Subsequently, we extend the treatment to more than one trace (as would be expected in dealing with donor and acceptor channels in smFRET). The method may be straightforwardly extended to the general case of more than two traces arising, for instance, with some single molecule
\Blue{fluorescent tweezers} experiments~\cite{neuman2008single,kellermayer1997folding,comstock2015direct} and experiments involving multiple FRET probes \cite{lee2007three}.

\subsection{Single time trace}
\label{sec:drift}

\Blue{We first consider a single experimental trace with measurements $\bar z=(z_1,z_2,\dots,z_N)$. 
At each time step, the measurements $z_n=x_n+y_n$ consist of the biomolecular emissions $x_n$ and drift $y_n$. Unlike emissions, which may abruptly change from time step to time step, the drift contribution to the time series, $\bar y=(y_1,y_2,\dots,y_N)$, 
evolves slowly. That is, drift evolves at a rate that is much slower than data acquisition. Assuming that the time scales at which the biomolecule and drift evolve are well separated, we may model each $y_n=f(t_n)$ as a point from a continuous function $f(t)$. Further, we may use the data to infer the shape of $f(t)$ within the Bayesian paradigm. The power of the method we propose here relies on two features: i) its generality which makes no assumptions about $f(t)$ besides continuity; and ii) the simultaneous learning of $f(t)$ (see below) with the rest of the model parameters as opposed to de-trending in pre-processing.}

\begin{figure}[t]
   \centering
   \includegraphics[scale=1]{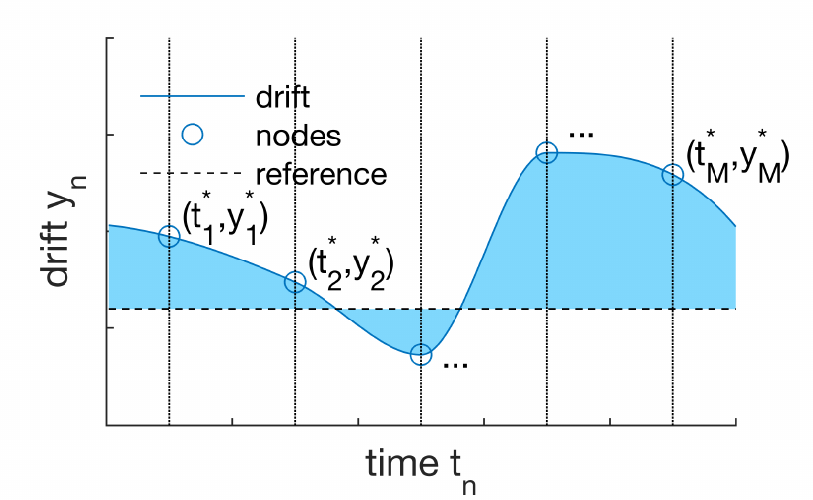} 
   \caption{{\bf Drift representation for a single experimental trace}. Circles mark the interpolation nodes $(t_m^*,y_m^*)$, the solid line marks the drift contribution to the time trace $\bar y=(y_1,y_2,\dots,y_N)$ and the dashed line marks the reference level $y_{\rm ref}^*$ which is set to the zero emission level. The shaded region represents the area $v^*$. For details see main text.}
   \label{fig:drift}
\end{figure}

\Blue{Although we intend to learn an arbitrary function $f(t)$, for computational reasons we use a finite approximation that is provided by interpolation. Given that interpolation provides approximations to any continuous function up to arbitrary degree \cite{atkinson2005theoretical}, this choice offers a 
flexible means of performing inference on the drift that is amenable to computation.}

\Blue{More precisely, to model the drift function $f(t)$, we apply a set of nodes $(t_m^*,y_m^*)$, $m=1,\dots,M$, and use \Blue{cubic spline} interpolation to approximate it across nodes, see Fig.~\ref{fig:drift}. For sufficiently large $M$, the resulting $f(t)$ are practically indistinguishable \cite{atkinson2005theoretical}. Therefore the precise choice of $M$ is of little or no importance with respect to biophysical applications. Concerning the placement of the nodes, we note that $t^*_m$ generally need not coincide with $t_n$. We also note that neither the interpolation needs to use splines as we will focus on here, but more general schemes can also be incorporated. For simplicity, in this study we assume that, once $M$ is chosen, $t_m^*$ are kept fixed and equidistantly placed over the time series. In contrast, the node heights $y_m^*$ are treated as parameters with values that are to be inferred from the data.}

In particular, to infer the values of $y_m^*$ which ultimately determine the overall shape of $f(t)$, we place a normal prior
\begin{linenomath*}\begin{align}
y_m^*\big| h^*,w^*&\sim\mathcal{N}(h^*,w^*).\label{eq:drift}
\end{align}\end{linenomath*}
Here, $h^*$ denotes the mean of the drift nodes and $w^*$ the precision. That is, $h^*$ influences the height at which nodes are placed and $w^*$ how tightly the nodes are placed around this average. To infer both $h^*$ and $w^*$, we specify a vague normal-gamma hyperprior.

The emission and drift models, resulting from Eqns.~\eqref{eq:emiss} and \eqref{eq:drift}, without further specifications are ill-posed. 
Given that contributions to the total emission from $\bar x$ and $\bar y$ are assessed only through the observations $\bar z=\bar x+\bar y$, the whole model is invariant 
with respect to a translation of  $\bar x$ and $\bar y$ equal in magnitude but opposite in direction. More precisely, the model gives identical outcomes for $\bar x+c$ and $\bar y-c$, for any arbitrary constant $c$. To resolve this issue, we may tether $\bar y$. That is, we may restrict the interpolation nodes $y_m^*$ such that the area between the predicted $\bar y$ and a reference level $y^*_{\rm ref}$, which is given by $v^*=\sum_{n=1}^N(y_n-y^*_{\rm ref})(t_{n}-t_{n-1})$, is zero. Thus, in our formulation, the interpolation nodes $y_m^*$ would not only satisfy the normal prior of Eq.~\eqref{eq:drift} but also satisfy the condition $v^*|y_m^*=0$. Finally, by setting $y_{\rm ref}^*=0$, we ensure that the drift trace remains near the zero emission level or equivalently we ensure that the estimated emission trace $\bar x$ remains close -- in fact, the closest possible -- to the observation trace $\bar z$.

In summary, the drift model we describe above assumes that: i) the drift stems from a continuous process; ii) its shape is dictated by the data; and iii) it remains close to the level of zero emissions. Under those assumptions, we can readily model any type of drift such as monotonic or oscillatory drift, such as in Figs.~\ref{fig:syn_1} and \ref{fig:exp_1} \Blue{that follow}, as the precise locations of $y^*_m$ adapt to the supplied data.

The full set of equations describing \icon, including the equations of iHMM and the drift representation, is summarized in the supporting materials.

\subsection{Multiple time traces}
\label{sec:multiple}

To handle observations collected by the same biomolecule contained in more than one parallel traces, we must extend the methodology presented above. 
Specifically, while the state transitions are coupled to all traces, we cannot assume that emission or drift models are coupled and must therefore learn these independently. Here we describe the necessary modifications. 

For simplicity, we assume that only two traces $\bar z^1=(z_1^1,z_2^1,\dots,z_N^1)$ and $\bar z^2=(z_1^2,z_2^2,\dots,z_N^2)$ are available. The general case of more than two traces is a straightforward extension.

As with the case of the single trace, observations are decomposed into emissions and drift $\bar z^1=\bar x^1+\bar y^1$ and $\bar z^2=\bar x^2+\bar y^2$. To model each emission trace, we assume distributions $F_{\sigma_k}^1(x^1)$ and $F_{\sigma_k}^2(x^2)$ different for each trace
\begin{linenomath*}\begin{align}
x^1_n\big| s_n&\sim F^1_{s_n},
\\
x^2_n\big| s_n&\sim F^2_{s_n},
\end{align}\end{linenomath*}
where $s_n$ denotes the conformation of the biomolecule at time $t_n$ which is modeled as in the single trace case. 
To obtain, $F_{\sigma_k}^1(x^1)$ and $F_{\sigma_k}^2(x^2)$, we use the same generic family, Eq.~\eqref{eq:F_emiss}; however, we apply different state specific parameters $\phi_{\sigma_k}^1$ and $\phi^2_{\sigma_k}$ on them. Furthermore, we use different prior distributions $H^1(\phi^1_\sigma)$ and $H^2(\phi^2_\sigma)$ for the parameters associated with each trace.

Similar to the case of the single trace, we model drift $\bar y^1$ and $\bar y^2$ using interpolation, but we place different sets of nodes $(t^1_m,y_m^{*1})$ and  $(t^2_m,y_m^{*2})$ on them with unique priors for each
\begin{linenomath*}\begin{align}
y_m^{*1}\big|h^{*1},w^{*1}&\sim\mathcal{N}(h^{*1},w^{*1}),
\\
y_m^{*2}\big|h^{*2},w^{*2}&\sim\mathcal{N}(h^{*2},w^{*2}),
\end{align}\end{linenomath*}
in addition to unique hyper-priors over $h^{*1},w^{*1}$ and $h^{*2},w^{*2}$. Finally, to deal with drift tethering, we restrict each drift trace to remain close to the zero emission levels individually $v^{*1}|y_m^{*1}=0$ and $v^{*2}|y_m^{*2}=0$. The full set of equations describing \icon for double trace analysis is summarized in the supporting materials.

\subsection{Computational considerations}

\Blue{For the analyses shown in the ~Results below}, we implemented \icon described above using a Gibbs sampling scheme \cite{robert2013monte}. Specifically, to deal with the infinite dimensional state space $\{\sigma_1,\sigma_2,\dots\}$ we used the beam sampling algorithm \cite{van2008beam} as described in our companion article~\cite{perspectives} that provides step-by-step implementation details for these methods. Briefly, the beam sampling algorithm uses slicer variables to truncate $\{\sigma_1,\sigma_2,\dots\}$ to a finite size that is adjusted during the sampling iterations~\cite{neal2000markov,walker2007sampling}. To account for drift, we combined a Metropolis random walk step within the overall Gibbs scheme \cite{robert2013monte}. 

More details are provided in our companion perspectives article~\cite{perspectives} and a working implementation of \icon, written in \texttt{MATLAB\textsuperscript{\textregistered}}, is provided in the supporting materials.\footnote{\Blue{This code can also be found on the authors' website as well as on GitHub.}}

The methodology developed here offers flexibility in the modeling and analysis of single molecule data. Nevertheless, this flexibility comes with added computational cost, a disadvantage inherited from Gibbs sampling \cite{robert2013monte}. In particular, our algorithm spends most of its time updating the state sequence $(s_1,s_2,\dots,s_N)$. As explained in Ref.~\cite{van2008beam}, this is performed by the forward filtering-backward sampling algorithm \cite{carter1994gibbs} which has a complexity of $O(NK^2)$. That is, the most costly part of our implementation scales linearly with the length of the trace $N$ and quadratically with the number of states $K$ that are needed to explain the observations. Despite this rather high complexity, a typical time trace from smFRET with 500--1000 data points, such as those analyzed below, take a reasonable time on an average desktop computer, $\approx$1~min, to sufficiently sample the posterior distribution.

We note that those components that make \icon differ from the simpler iHMM only trivially add to the computational burden. Rather, the bulk of the burden originates from the computational schemes required to characterize the posterior (i.e.\ sampling) of the iHMM itself. Computational Statistics is an active area of research with variational methods \cite{ding2010variational,johnson2014stochastic} \Blue{or approximate sampling schemes} \cite{tripuraneni2015particle} actively considered as alternatives to brute force Gibbs sampling. On this basis, we believe that this burden of the iHMM will be substantially reduced 
in the near future.

\section{Results}
\label{sec:results}

Below we present selected applications of the method whose mathematics are detailed in the previous section. 
Initially, to benchmark our method, we apply it to 
the analysis of synthetic data.
We subsequently tackle experimental smFRET traces provided by the Nils Walter lab 
available online (see the supporting materials of Ref. \cite{blanco2015single}). 
These traces had previously been analyzed 
by traditional HMMs~\cite{blanco2015single}.

\begin{figure}[t]
   \centering
   \includegraphics[scale=1]{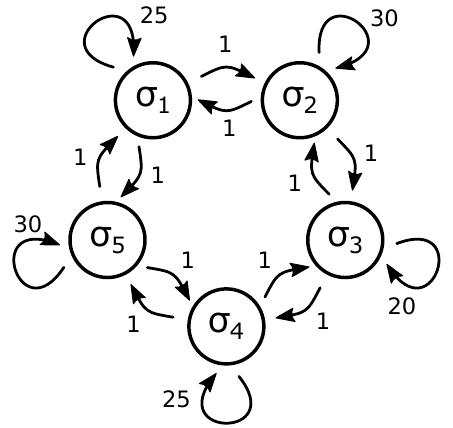} 
   \caption{\Blue{\bf Five-state Markov model generating the synthetic data that are used to benchmark \icon}. The model transitions between five states $\sigma_1$--$\sigma_5$ with transition probabilities proportional to the weights shown. For simplicity, only the state transitions shown (arrows) are allowed. The generated trace, after the addition of noise and drift, is shown on Fig.~\ref{fig:syn_1}.}
   \label{fig:syn_hmm}
\end{figure}

\subsection{Synthetic data}
\label{sec:syn}

\begin{figure}[tbp]
   \centering
   \includegraphics[scale=1]{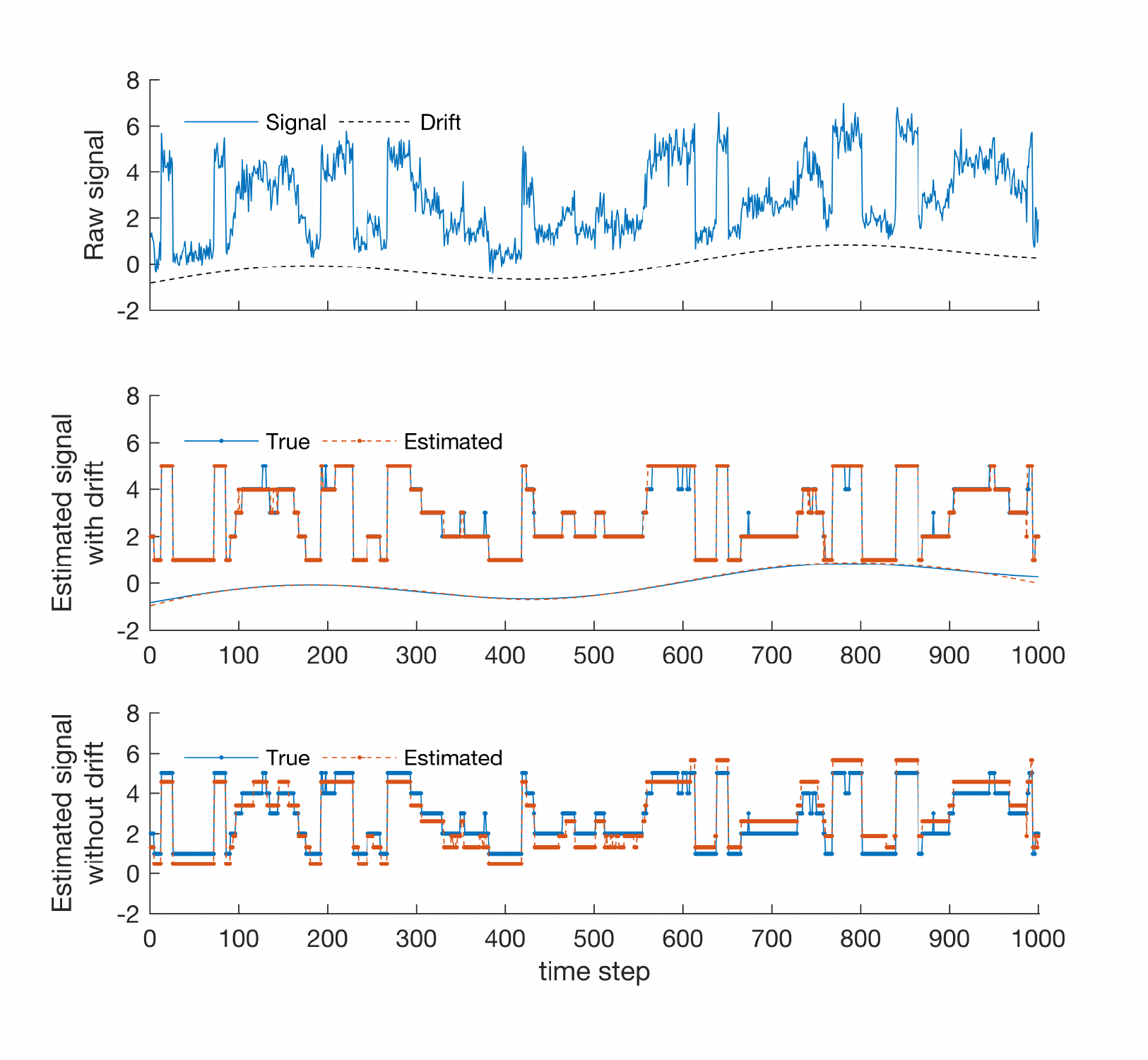} 
   \caption{{\bf \icon can analyze time traces contaminated with both noise and drift and self-consistently learn the drift while learning the number of states in addition to all other quantities determined by the HMM}.
\emph{Upper panel:} Synthetic time traces with drift were generated as described in~Fig.~\ref{fig:syn_hmm}.
\emph{Middle panel:} True (solid) and estimated (dashed) traces for the state sequence means and drift. \emph{Lower panel:} Corresponding true and estimated trace for the state sequence means without drift estimation. Note that without drift correction, the iHMM over-interprets drift as the population of additional states. This over-interpretation is further quantified in Fig.~\ref{fig:syn_2} (left panels).}
   \label{fig:syn_1}
\end{figure}

We use state transitions of the Markov model shown on Fig.~\ref{fig:syn_hmm} to generate the signal underlying the synthetic trace. As for the noise, we assumed Gaussian emissions of different mean values and precisions associated with each state. In addition to noise, we contaminated the resulting trace with drift  
containing a linear trend as well as oscillations. The resulting trace is shown on Fig.~\ref{fig:syn_1} (upper panel).
We choose five states with a reasonably high noise level, as can be seen in Fig.~\ref{fig:syn_1}, specifically because fixing the number of states or subtracting drift {\it a priori} for the HMM would be difficult.

The middle panel of  Fig.~\ref{fig:syn_1} shows  the estimated most likely state sequence and drift. 
A comparison with the true emission trace (the trace plotting the means of the emissions of the states visited) 
reveals good agreement with only few occasional missed transitions. The lower panel shows how state estimation badly 
deteriorates had we ignored the drift altogether.

\begin{figure}[t]
   \centering
   \includegraphics[scale=1]{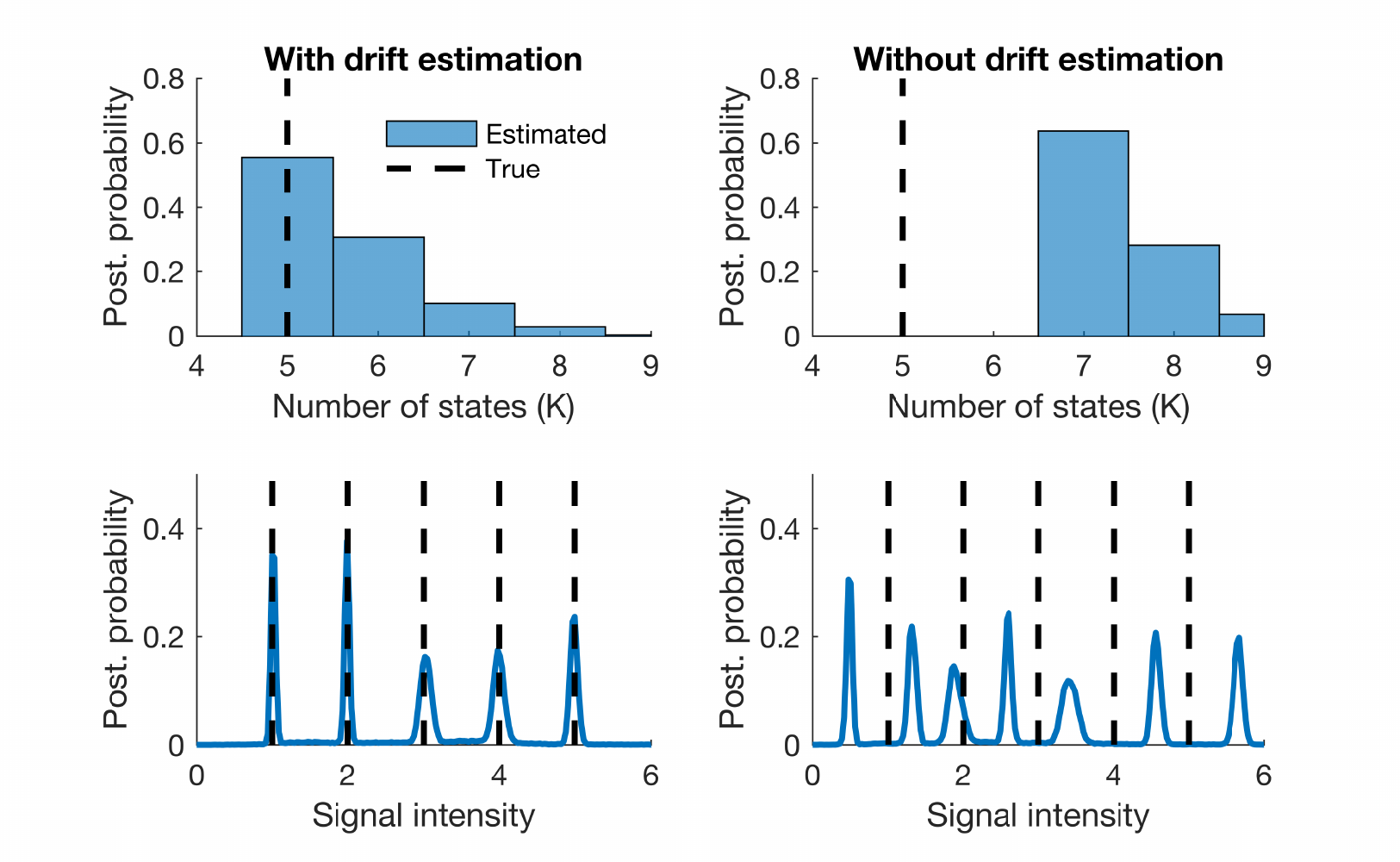} 
   \caption{{\bf \icon yields a full posterior distribution over states and their associated parameters}. 
   \emph{Upper panels:} Estimated number of states that generate the data in Fig.~\ref{fig:syn_1}.
    \emph{Lower panels:} Estimated locations of the emission mean levels. Both sets of results are shown for the \icon with drift estimation (left panels) and without (right panels). In all panels, dashed lines indicate the true values. The introduction of artifact states on the right panels is apparent which indicates the limits of the iHMM (i.e.~ignoring drift) in the analysis of time series.}
   \label{fig:syn_2}
\end{figure}

Unlike the HMM,  the iHMM -- from which \icon is derived --- builds a full posterior distribution 
over the number of states present in the trace as shown in Fig.~\ref{fig:syn_2}.
In particular, in Fig.~\ref{fig:syn_2} (upper left panel), we show a histogram of the number of states estimated by  
\icon.  As can be seen,  \icon's marginal posterior over the  number of states is peaked at the correct number, 5. Since the iHMM models a state space of size that is not 
fixed, this number may take different values. For example, \icon provides low probabilities to other choices, 6, 7, 8, and 9, indicating that it is less likely the supplied data would have been generated by a system with this many states. Cases with fewer that 5 or more than 9 states effectively have no probability. Furthermore, by ignoring drift, the corresponding estimates of Fig.~\ref{fig:syn_2} (upper right panel) shift toward higher numbers suggesting the recruitment of ``artifact states".

In addition, Fig.~\ref{fig:syn_2} (lower left panel) shows how the mean signal levels of each state are distributed. This figure clearly illustrates the locations of the state emissions which -- consistent with the upper panel -- identify 5 prominent peaks. As can be seen, \icon successfully identifies and localizes the emission distributions despite the large number of states and the presence of drift in the raw trace, Fig.~\ref{fig:syn_1} (upper panel). 

The right panels of Fig.~\ref{fig:syn_2} illustrate just how inaccurate the inference becomes if we ignore drift. 
Here a plain iHMM with no drift correction clearly overestimates the number of states and misidentifies the emission distributions in large part by over-interpreting the location of the mean levels.

\subsection{Experimental data}

In the analysis of real data, we focused on smFRET measurements. In typical smFRET experiments, two parallel traces are available describing the same system: the donor and acceptor trace~\cite{blanco2010analysis}.  We denote the  trace intensity levels recorded in the acceptor and donor channels as $\bar x^1$  and $\bar x^2$, respectively. 
Both traces are commonly combined into a single ``FRET efficiency" trace $\bar x=\bar x^1/(\bar x^1+\bar x^2)$.

For this study, we used selected experimental traces available from Ref.~\cite{blanco2015single} which describes details of the sample preparation and data acquisition. 
We analyzed these traces with \icon utilizing either a single trace, for which we used the efficiency $\bar x$, or both acceptor $\bar x^1$ and donor $\bar x^2$ recordings. The same traces have been previously analyzed using traditional HMM with the results of the analysis available in Ref.~\cite{blanco2015single}.

\subsubsection{FRET: single trace}

\begin{figure}[t]
   \centering
   \includegraphics[scale=1]{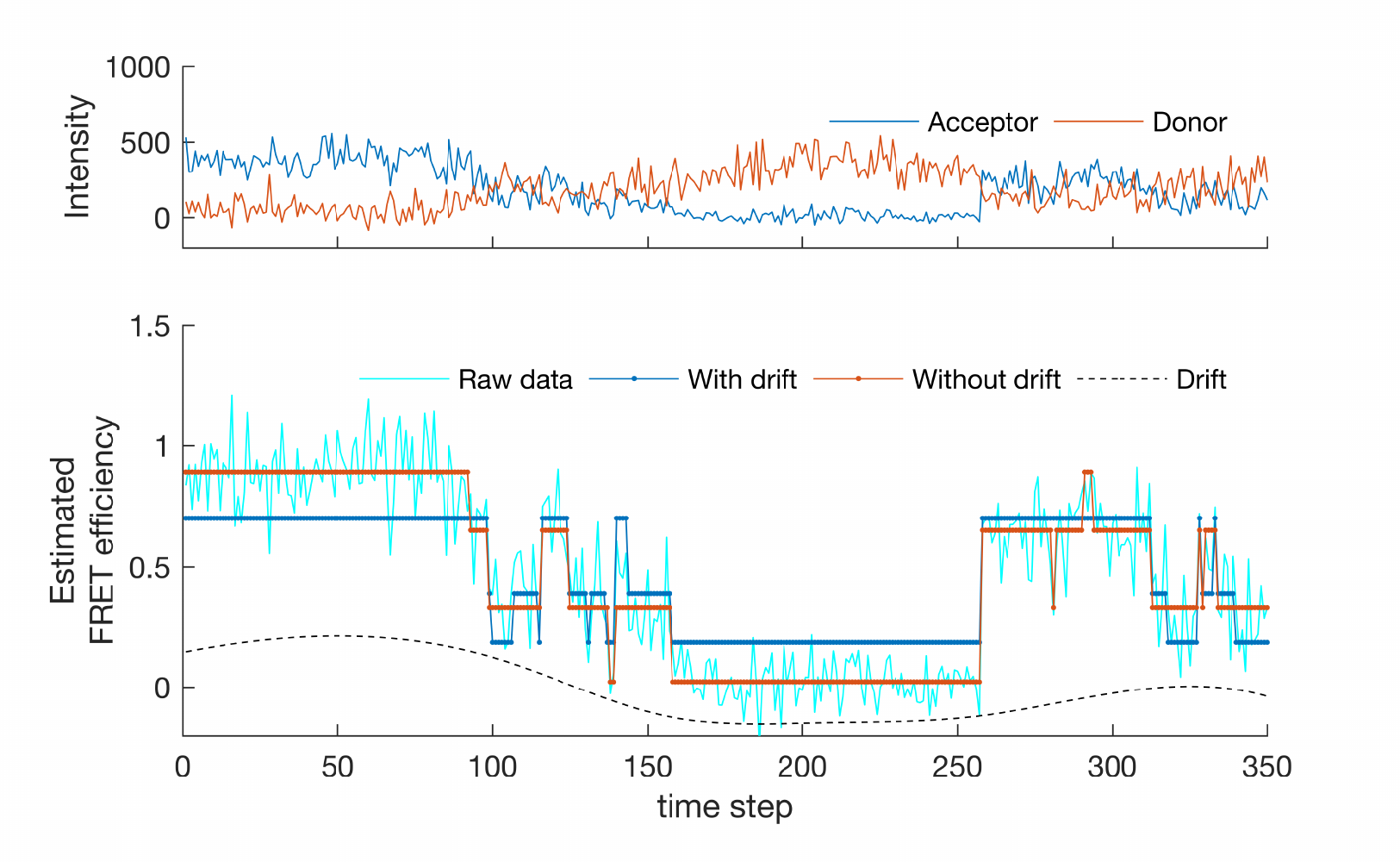} 
   \caption{{\bf \icon applied in the analysis of experimental smFRET time traces}. \emph{Upper panel:} Raw acceptor and donor traces from smFRET measurements~\cite{blanco2015single}. For the analysis, these traces were combined into a single FRET efficiency trace. \emph{Lower panel:} Corresponding raw and estimated FRET efficiency. Here we show the results that consider drift (blue and black lines) and results that ignore drift (red line). Note how ignoring drift leads to over-interpretation of the higher and lower efficiency levels. This over-interpretation is further quantified in Fig.~\ref{fig:exp_2}.}
   \label{fig:exp_1}
\end{figure}

\begin{figure}[t]
   \centering
   \includegraphics[scale=1]{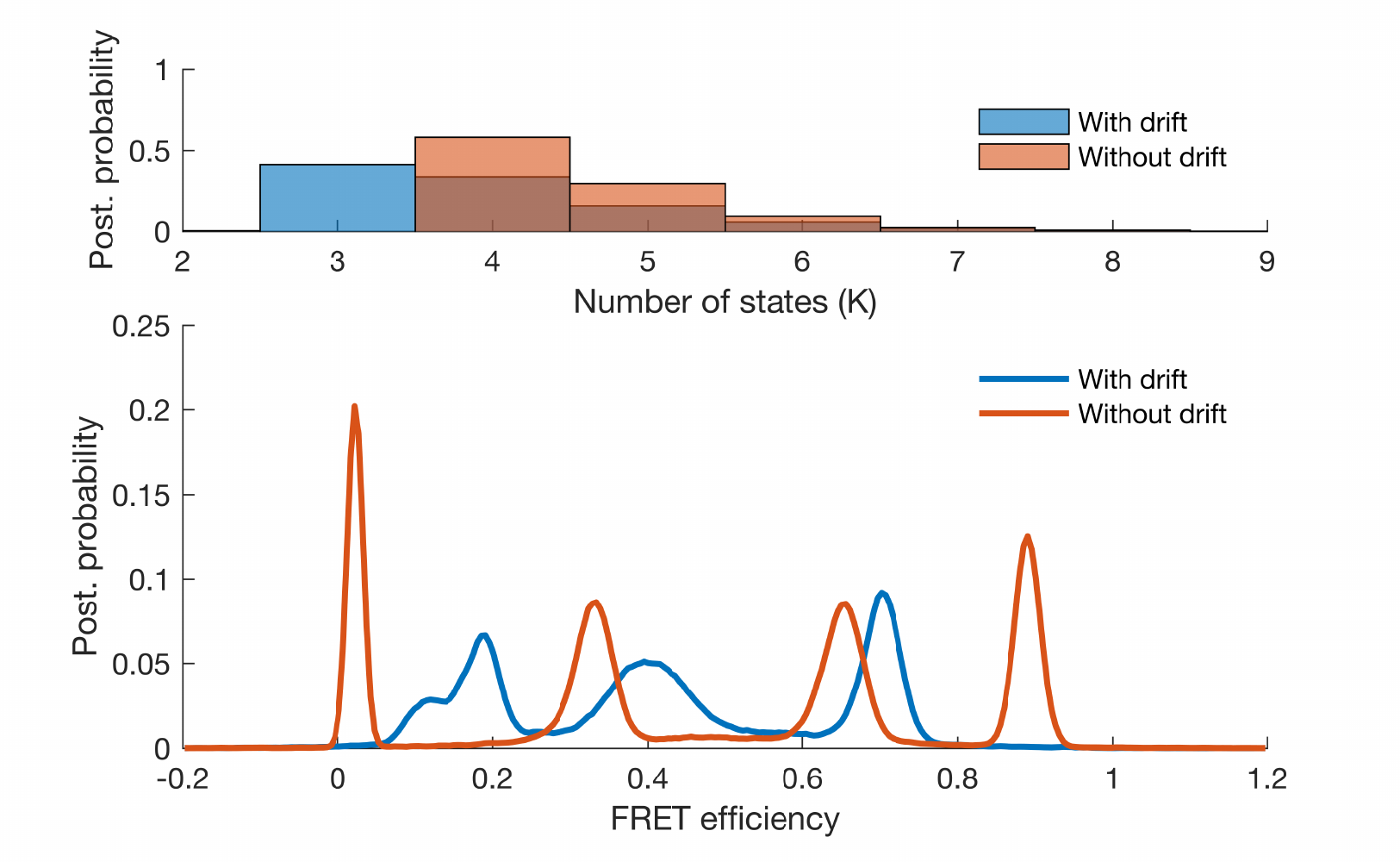} 
   \caption{{\bf Results of the \icon analysis on the experimental smFRET data of Fig.~\ref{fig:exp_1}}. \emph{Upper panel:} Estimated number of states. \emph{Lower panel:} Estimated locations of the emission distributions. In both panels we show results incorporating drift (blue) and ignoring drift (red). Over-interpretation of the data in the latter case is apparent.}
   \label{fig:exp_2}
\end{figure}

Figure~\ref{fig:exp_1} (upper panel) displays individual acceptor and donor intensity recordings. 
Figures~\ref{fig:exp_1} (lower panel) and \ref{fig:exp_2} show the resulting estimates provided by \icon. To obtain the estimates in these figures, 
we used \icon once assuming drift and once ignoring drift similar to the synthetic dataset of the previous section.

As can be seen, \icon with drift identifies 3 states located approximately at efficiencies 0.2, 0.4, and 0.7. As with the synthetic traces, ignoring the drift, the iHMM over-interprets the data and identifies 4 states instead located approximately at 0, 0.35, 0.65, and 0.9. The latter estimates are similar to the estimates provided in Ref.~\cite{blanco2015single} obtained by means of traditional HMMs, which i) ignore drift and ii) fix the total number of states \emph{a priori}. 

A few points are in order here on the problems caused by drift for even the simple traditional HMM. 
First, drift in a time trace necessarily results in a closer fit of the model to the data when an initial larger number of states are introduced. 
Thus, without appropriately accounting for drift, a different number of states used to model the data across data sets 
may simply reflect the level of drift.
Furthermore, regardless of the numbers of states used in an HMM analysis (even if a reasonable number of states are fixed {\it a priori}), 
the values for the FRET efficiencies -- that ultimately are interpreted as inter-molecular distances -- are off in the absence of drift correction.

\subsubsection{FRET: multiple traces}

\begin{figure}[t]
   \centering
   \includegraphics[scale=1]{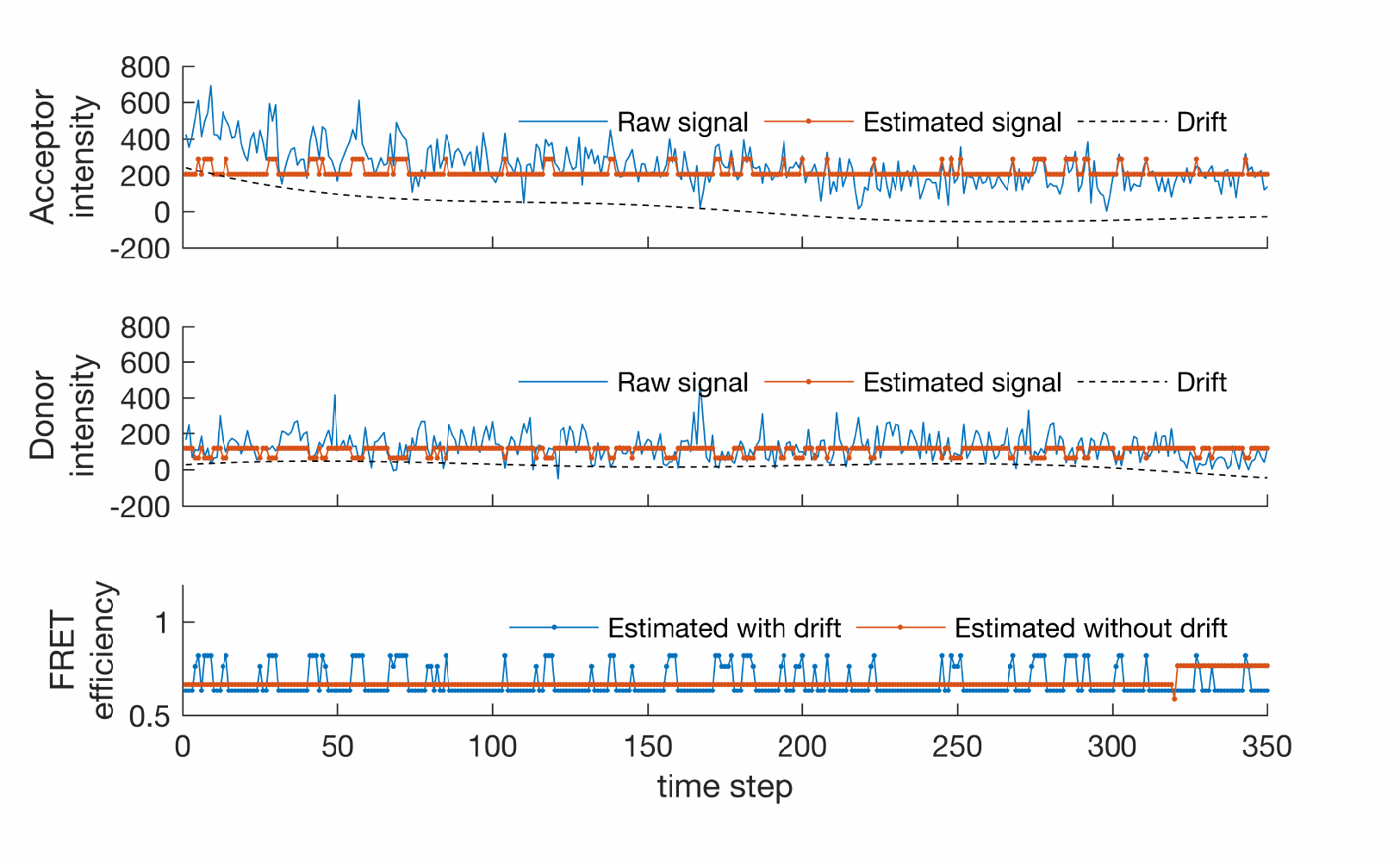} 
   \caption{{\bf Example of smFRET donor and acceptor traces analyzed individually by \icon.} \emph{Upper panel:} Acceptor raw recordings and estimated signal. \emph{Middle panel:} Donor raw recordings and estimated signal. \emph{Lower panels:} Corresponding estimated FRET efficiency with drift and without drift estimation. Note the homogenization of the signal induced by the drifted acceptor trace in the latter case.}
   \label{fig:double_1}
\end{figure}

Figure~\ref{fig:double_1} (upper two panels) shows an example of acceptor and donor traces where the acceptor trace drifts heavily upward toward its beginning. We applied \icon on these traces, but instead of using the resulting FRET efficiency we utilized the recordings individually. As demonstrated in Fig.~\ref{fig:double_1} (lower panel), our method identifies 3 states located at efficiencies of approximately 0.65, 0.75, and 0.80 as compared to the estimate of 0.55 and 0.75 (not shown) using a finite HMM where two states had previously been \emph{a priori} imposed \cite{blanco2015single}.
The difference in the efficiencies here is primarily ascribed to the drift that we corrected for in our method.

Again, to demonstrate the effects of ignoring drift, we analyzed the same traces disabling drift correction. In this case, the estimated number of states remains 3 and their locations are at 0.60, 0.65 and 0.75. Nevertheless, despite the apparent similarities, there is dramatic difference in the kinetics. For example, the lower panel compares the most likely signal sequence of the two cases. Due to the heavily drifted acceptor trace, essentially the whole trace is estimated to consist of  two excessively long dwells and only a short transition to a third state. 

Previously we had seen how drift may force the iHMM to populate artifact states. But in fact, drift can also do the opposite.
As can be seen, with \icon, the biomolecule in Fig.~\ref{fig:double_1} is estimated to alternate quickly between states for all times suggesting fast kinetics and shorter dwell times. By contrast, by ignoring drift the biomolecule is estimated to dwell in a single state throughout the entire first half of the trace, suggesting slow kinetics and a long dwell time for this state. In other words, the heavily drifting acceptor trace essentially homogenizes the data and makes transitions between states \Blue{more difficult to detect}.

\section{Discussion}
\label{sec:discussion}

The HMM has been a workhorse of time series analysis in single molecule Biophysics. Having started with the analysis of time series derived from  
single ion-channel patch clamp experiments~\cite{qin2000direct,venkataramanan2002applying,chung1990characterization}, HMMs have then been used
in the analysis of smFRET~\cite{blanco2010analysis} and single molecule force spectroscopy \cite{kruithof2009hidden}, amongst other methods \cite{tavakoli2016single}. 
The HMMs greatest shortcoming -- one that has been elegantly resolved by the iHMM \cite{beal2001infinite} -- 
is that the number of states attained need to be prespecified {\it a priori}. The challenge remains that, as is, the power of the 
iHMM cannot be directly harnessed to tackle single molecule problems.

The method presented here, \icon, generalizes the iHMM to deal with single molecule time traces directly
by coupling the iHMM to a slow-evolving control process that accounts for drift. We also described how one may apply \icon
to problems where multiple traces are provided on a single system like in smFRET.

Throughout this study we dealt  with cases of drift evolving on a time scale slower than typical biomolecular dynamics, for example as in Fig.~\ref{fig:syn_1}. However, this is not a limitation of the method. In fact, while less biophysically relevant perhaps, the method can also be applied with modifications to handle non-smooth drift or even drift that changes at time scales as fast as the  dynamics themselves. This can be achieved by increasing the number of nodes in the interpolation, see Fig.~\ref{fig:drift}, or by using a different set of basis functions in the interpolation of $f(t)$. For example, cubic polynomials that are used in the analyses shown, can be replaced by Hermite polynomials or step functions
and  \Blue{future work on these underdetermined problems will be dictated by their experimental need.}

A few other methods outside Biophysics have considered continuous processes coupled to iHMMs as in the present study. These include the switching linear dynamical system (SLDS) \cite{fox2009nonparametric} and its variants \cite{fox2011bayesian} which couple the dynamics of an array of linear systems with those of an iHMM. Specifically, SLDS uses linear dynamics to model traces of observations that evolve in a continuous manner which is best suited to describe mechanical systems. The formulation we adopted here was inspired by these methods but moves in a different direction as our goal is not to provide an accurate description of a system with well known dynamics but rather to provide a general modeling approach that relies on minimal assumptions about the time evolution of the drift.
A more general method with less restrictive requirements is indeed demanded by 
single molecule experiments as assumptions on the drift's dynamics are difficult to make {\it a priori}  \cite{roy2008practical,holden2010defining}.

As we show in the previous section, ignoring drift can lead to dramatic differences in the outcome of the analysis. 
Most notably, in all cases examined -- synthetic or real -- drift forces the iHMM to introduce artifact states 
that give rise to erroneous kinetics or transitions between real and artifact states. 
With heavily drifting time series, these effects are exaggerated.

\Blue{Drift correction in \icon} fundamentally avoids the cherry-picking of raw time traces or pre-processing of traces that would otherwise occur when presented with traces that drift to varying degrees. \Blue{To date, heavily drifted time series are either discarded from the analysis based on subjective criteria or corrected for drift in ad hoc and often undocumented manner.} Indeed, as we have demonstrated, the analysis of traces with drift under the appropriate correction leads to accurate results and thus there is no need to discard traces on an arbitrary basis. \Blue{What is more, state determination in \icon also avoids post-processing model selection steps to determine the number of states using tools such as maximum evidence as for example in Ref.~\cite{bronson2009learning}.}

Finally, to account for multiple experimental traces, we coupled the emissions to several observation traces to specifically deal with data such as smFRET. In doing so, we fitted separate emission distributions and drift in each trace individually. The method presented here is general in that it assumes no correlation in the different traces. However, in many application there might be additional structure shared between observations besides what we have represented in the present study. For example, in FRET measurements, donor and acceptor emissions are strongly anti-correlated \cite{roy2008practical}. In such cases, incorporating further assumptions on the model, for example by explicitly forcing the model to fit anti-correlated emission distributions, could lead to improved inference. Such modifications are readily accommodated within \icon.

It took roughly thirty years for the theoretical breakthrough -- that is
Bayesian nonparametrics \cite{ferguson1983bayesian} and the Dirichlet process \cite{teh2012hierarchical} --
to become computationally tractable \Blue{and have a deep impact} in data science. 
With powerful computational tools at hand and many more active researchers in the field, it is our expectation that
it will take a fraction of that time for nonparametrics to make an equally important contribution to single molecule Biophysics.





\singlespacing

\bibliographystyle{unsrt}
\bibliography{ioannis_ref_ihmm,ioannis_ref_ihmm_ext}

\end{document}